\documentstyle[prb,aps,epsfig]{revtex}

\begin{document}

\draft

\title{Diversity of discrete breathers observed in a Josephson ladder}

\author{P. Binder, D. Abraimov, and A. V. Ustinov}
\address{Physikalisches Institut III, Universit\"at Erlangen-N\"urenberg,
  Erwin-Rommel-Stra\ss e 1, D-91058 Erlangen, Germany}

\date{\today}

\wideabs{

\maketitle

\begin{abstract}
  We generate and observe discrete rotobreathers in Josephson
  junction ladders with open boundaries. 
  Rotobreathers are localized excitations that persist under the
  action of a spatially uniform force. We find a rich variety of
  stable dynamic states including pure symmetric, pure asymmetric, and
  mixed states.  The parameter range where the discrete breathers are
  observed in our experiment is limited by retrapping due to
  dissipation.
\end{abstract}



}

Nonlinearity and lattice discreteness lead to a generic class of
excitations that are spatially localized on the scale comparable to
the lattice constant. These excitations, also known as {\em discrete
  breathers}, have recently attracted a lot of interest in theory of
nonlinear lattices \cite{sa97,sfcrw98}. It is believed that discrete
breathers might play an important role in the dynamics of various
physical systems consisting of coupled nonlinear oscillators. It has
been even said that that for discrete nonlinear systems breathers
might be as important as solitons are for the continuous nonlinear
media.

There have been several recent experiments that reported on generation
and detection of discrete breathers in diverse systems. These are
low-dimensional crystals \cite{Swanson99}, anti-ferromagnetic
materials \cite{Schwarz99}, coupled optical waveguides \cite{hseys98},
and Josephson junction arrays \cite{etjjmtpo99,binder00}.  By using
the method of low temperature scanning laser microscopy, we have
recently reported the first direct visualization of discrete breathers
\cite{binder00}.

In this paper we present new measurements of localized modes in
Josephson ladders. Using the same method as in our first experiment,
we study an even more tightly coupled lattice of Josephson junctions
and observe a rich diversity of localized excitations that
persist under the action of a spatially uniform force.

A biased Josephson junction behaves very similar to its mechanical
analog that is a forced and damped pendulum.  An electric bias current
flowing across the junction is analogous to a torque applied to the
pendulum. The maximum torque that the pendulum can sustain and remain
static corresponds to the critical current $I_{\rm c}$ of the
junction. For low damping and bias below $I_{\rm c}$, the junction
allows for two states: the superconducting (static) state and the
resistive (rotating) state.  The phase difference $\varphi$ of the
macroscopic wave functions of the superconducting islands on both
sides of the junction plays the role of an angle coordinate of the
pendulum.  According to the Josephson relation, a junction in a
rotating state generates dc voltage $V=\frac{1}{2\pi}\Phi_0 \left<
  \frac{d\varphi}{dt} \right>$, where $\left<...\right>$ is the time
average.  By connecting many Josephson junctions by superconducting
leads one gets an array of coupled nonlinear oscillators.

We perform experiments with a particular type of Josephson junction
array called Josephson ladder. Theoretical studies
\cite{fmmfa96,sfms99,Mazo99} of these systems have predicted the
existence of spatially localized excitations called rotobreathers. 
Rotobreathers are $2\pi$-periodic solutions in time that are
exponentially localized in space.

\begin{figure}[htb]
\vspace{5pt}
\centerline{\epsfig{file=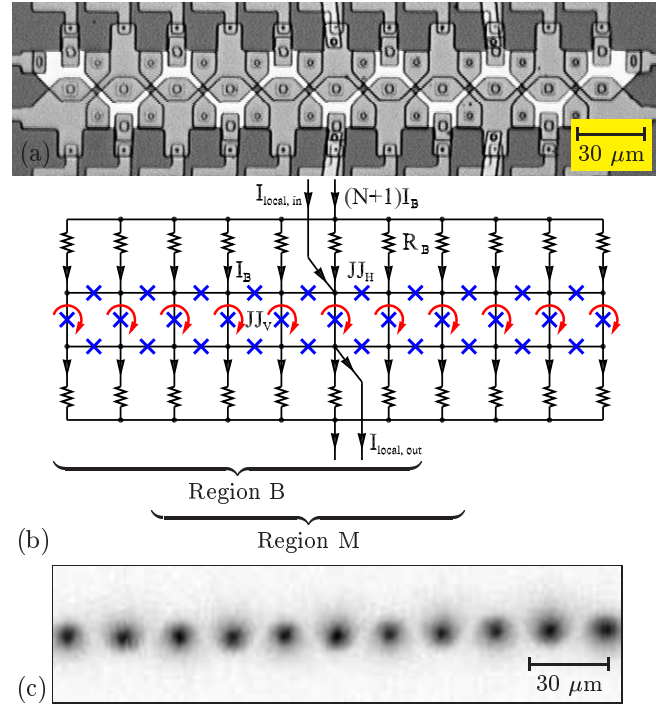,width=3.4in}}
\vspace{5pt}
\caption{Optical (a) and schematic (b) view of a linear ladder.  (c)
  Spatially homogeneous whirling state measured by the laser scanning
  technique.}
\label{schematic}
\end{figure}

Measurements are performed on linear ladders (with open boundaries)
consisting of Nb/Al-AlO$_x$/Nb underdamped Josephson tunnel junctions
\cite{Hypres}.  An optical image Fig.~\ref{schematic}(a) shows a
linear ladder that is schematically sketched in
Fig.~\ref{schematic}(b). Each
cell contains 4 small Josephson junctions.  The size of the hole
between the superconducting electrodes which form the cell is about
$3\times 3$ $\mu$m$^2$. The distance between the Josephson junctions
is about $24$ $\mu$m.  The number of cells $N$ in the ladder
is 10.  Measurements presented in this paper have been performed at
$5.2$ K.  The bias current
$I_{\rm B}$ was uniformly injected at every node via thin-film
resistors $R_{\rm B}=32$ $\Omega$.  Here we define {\sl vertical}
junctions (${\sl JJ_V}$) as those in the direction of the external
bias current, and {\sl horizontal} junctions (${\sl JJ_H}$) as those
transverse to the bias.  The ladder voltage was read across the
central vertical junction.  The damping coefficient
$\alpha=\sqrt{\Phi_0/(2\pi I_{\rm c}CR_{\rm sg}^2)}$ is the same for
all junctions as their capacitance $C$ and sub-gap resistance $R_{\rm
  sg}$ scale with the area and $C_{\rm H}/C_{\rm V}=R_{\rm sgV}/R_{\rm
  sgH}$.  The damping $\alpha$ in the experiment can be controlled by
temperature and its typical values vary between $0.1$ and $0.02$.

There are two types of coupling between cells in a ladder.  The first
is the inductive coupling between the cells that is expressed by the
self-inductance parameter $\beta_{\rm L} = 2\pi L I_{\rm cV}/\Phi_0$,
where $L$ is the self-inductance of the elementary cell.  The second
is the nonlinear Josephson coupling via horizontal junctions.  The
ratio of the horizontal and vertical junction areas is called the
anisotropy factor and can be expressed in terms of the junction
critical currents $\eta = I_{\rm cH}/I_{\rm cV}$.  If this factor is
equal to zero, the vertical junctions are decoupled and operate
independently one from another.  On the other hand, if this factor
goes to infinity the ladder behaves like a parallel one-dimensional
array and no rotobreather can exist since no magnetic flux can enter
through the horizontal junctions. Thus, it is an important challenge
to increase the anisotropy as far as possible.  In this work we
present measurements for the highest anisotropy factor ($\eta = 0.56$)
studied up to now. In contradiction to the existing theoretical
prediction \cite{Mazo99} for this anisotropy value, in the studied
parameter range with moderate dissipation we find a rich diversity of
localized excitations.

To briefly introduce the role of parameters we like to quote the
equations of motion for our system (see Ref.~\onlinecite{sfms99} for
details):
\begin{eqnarray}
&\ddot{\varphi}^{\rm V}_l + \alpha \dot{\varphi}^{\rm V}_l + 
\sin \varphi^{\rm V}_l =
\gamma
-\frac{1}{\beta_{\rm L}}(-\Delta \varphi^{\rm V}_l + \nabla
\tilde{\varphi}^{\rm H}_{l-1}
- \nabla \varphi^{\rm H}_{l-1}),
\label{1-1} \\
&\ddot{\varphi}^{\rm H}_l + \alpha \dot{\varphi}^{\rm H}_l + 
\sin \varphi^{\rm H}_l =
-\frac{1}{\eta \beta_{\rm L}}(\varphi^{\rm H}_l - \tilde{\varphi}^{\rm
H}_l +
\nabla \varphi^{\rm V}_l),
\label{1-2} \\
&\ddot{\tilde{\varphi}}^{\rm H}_l + \alpha \dot{\tilde{\varphi}}^{\rm
H}_l + 
\sin \tilde{\varphi}^{\rm H}_l =
\frac{1}{\eta \beta_{\rm L}}(\varphi^{\rm H}_l - \tilde{\varphi}^{\rm
H}_l +
\nabla \varphi^{\rm V}_l),
\label{1-3}
\end{eqnarray}
where $\varphi^{\rm V}_l, \varphi^{\rm H}_l, \tilde{\varphi}^{\rm
  H}_l$ are the phase differences across the $l$th vertical junction
and its right upper and lower horizontal neighbors, $\nabla \varphi_l
= \varphi_{l+1} - \varphi_l$, $\Delta \varphi_l = \varphi_{l+1} +
\varphi_{l-1} - 2\varphi_l$, and $\gamma = I_{\rm B} / I_{\rm cV}$ is
  the normalized bias current.

In order to generate a discrete breather in a ladder we used the
technique described in Ref.~\onlinecite{binder00}.  We used two extra
bias leads for the middle vertical Josephson junction (cf.\ 
Fig.~\ref{schematic}(b)) to apply a local current $I_{\rm local} >
I_{\rm
  cV}$.  This current switches the vertical junction in the resistive
state.  At the same time, being forced by magnetic flux conservation,
one or two horizontal junctions on both sides of the vertical
junctions also switch to the resistive state.  After that $I_{\rm
  local}$ is reduced and, simultaneously, the uniform bias $I_{\rm B}$
is tuned up.  In the final state, we keep the bias $I_{\rm B}$ smaller
than $I_{\rm cV}$ and have reduced $I_{\rm local}$ to {\em zero}.  By
changing the starting value of $I_{\rm local}$ it is possible to get
more than one vertical junction rotating. When trying the slightly
different current sweep sequence described in
Ref.~\onlinecite{etjjmtpo99}
we generated mainly states with many rotating vertical junctions.

In all measurements presented below, we have measured the dc voltage
across the middle vertical junction as a function of the homogeneous
bias current $I_{\rm B}$.  We used the method of low temperature
scanning laser microscopy \cite{LTSLM} to obtain electrical images
from the dynamic states of the ladder.  The laser beam locally heats
the sample and changes the dissipation in an area of several
micrometers in diameter.  If the junction at the heated spot is in the
resistive state, a voltage change will be measured.  By scanning the
laser beam over the whole ladder we can visualize the rotating
junctions.

\begin{figure}[htb]
\vspace{5pt}
\centerline{\epsfig{file=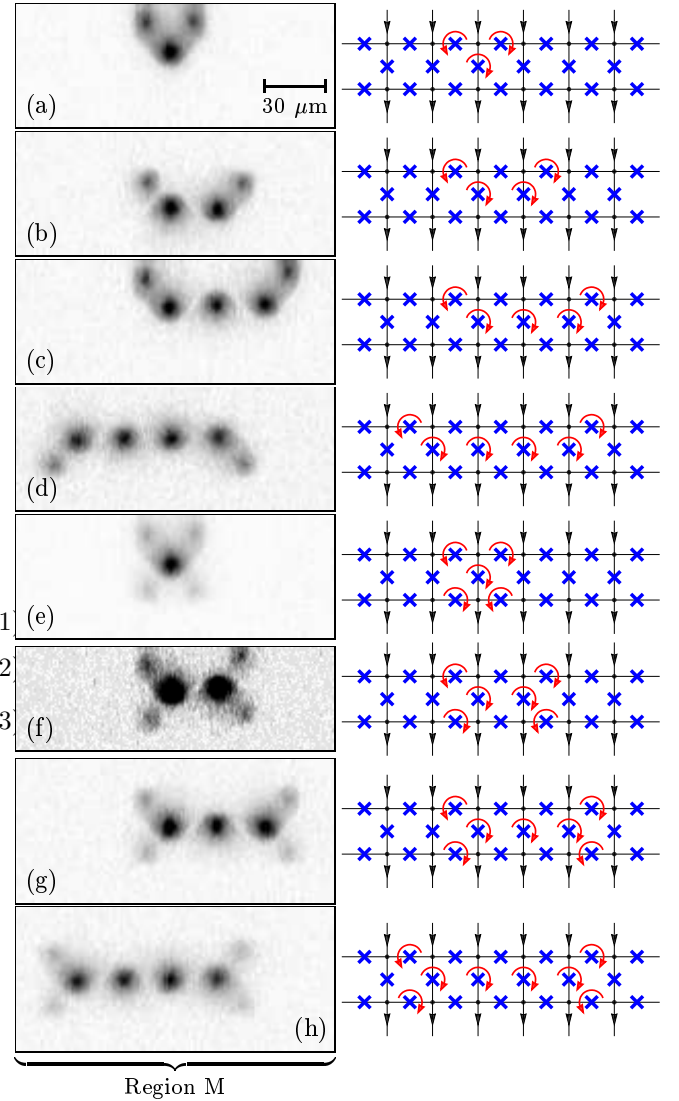,width=3.4in}}
\vspace{5pt}
\caption{
  Various localized states (discrete rotobreathers) measured by the
  low temperature scanning laser microscope: (a)--(d) asymmetric
  rotobreathers, (e)--(h) symmetric rotobreathers.  Region M is
  illustrated in Fig.~\ref{schematic}(b).}
\label{breather}
\end{figure}

\begin{figure}[htb]
\vspace{5pt}
\centerline{\epsfig{file=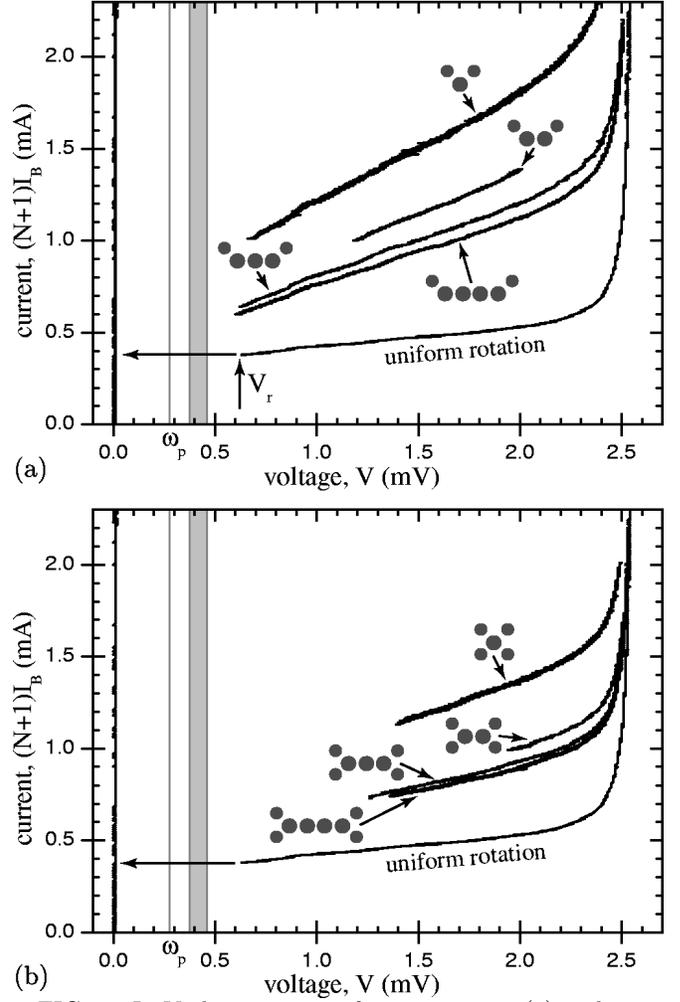,width=3.4in}}
\caption{$I_{\rm B}$-$V$ characteristics for asymmetric (a) and
  symmetric (b) breathers in a ladder with the parameters
  $N=10$, $I_{\rm cV}=320$ $\mu$A, $\eta=0.56$, and $\beta_{\rm
    L}=4.3$.  The grey region indicates the frequency range of the
  upper plasmon band.  $\omega_{\rm p}$ is the plasma frequency.}
\label{i-v}
\end{figure}

Various measured ladder states are shown in Fig.~\ref{breather} as 2D
gray scale maps.  The gray scale corresponds to the measured voltage
response during the laser scanning. In Fig.~\ref{breather}(a) we
present the simplest of observed states, an {\em asymmetric\/}
single-site rotobreather, where one vertical Josephson junction and
the upper adjacent horizontal Josephson junctions are in the resistive
state.  On the right side of the plot we show the corresponding
schematic view with arrows marking the rotating junctions. In the case
of Fig.~\ref{breather}(b) two vertical junctions and two horizontal
junctions are rotating that corresponds to an asymmetric two-site
breather.  An asymmetric three-site breather and an asymmetric
four-site breather are shown in Fig.~\ref{breather}(c) and (d)
respectively. Here, the voltage of the whirling horizontal and
vertical junctions are equal due to magnetic flux conservation. The
magnetic flux enters the ladder through one horizontal junction, goes
through the vertical ones and leaves the ladder through another
horizontal junction. When a single magnetic flux quantum is passing
through a Josephson junction its phase $\varphi$ is changing by
$2\pi$. In contrast to our previous measurements of an annular ladder
\cite{binder00},here in ladders with open boundaries we observe
asymmetric breathers as frequently as symmetric breathers.  The
$I_{\rm B}$-$V$ characteristics of these localized states is presented
in Fig.~\ref{i-v}(a). Particular states indicated on the plot are
stable along the measured curves. The more junctions are whirling, the
higher is the measured resistance.

Another type of discrete breathers observed in our experiment is shown
in the Fig.~\ref{breather}(e)--(h).  Figure~\ref{breather}(e)
illustrates a {\em symmetric\/} single-site breather with one vertical
and all four adjacent horizontal junctions are in the resistive state.
Though we call this states symmetric, the upper and the lower
horizontal junctions may in general have different voltages.  The
simplest voltage-symmetric case would be when each horizontal junction
voltage is equal to half of the voltage of the whirling vertical
junction \cite{binder00}.  We also observed a variety of multi-site
breathers of this type.  A two-site, a three-site, and a four-site
breather is shown in Fig.~\ref{breather} (f), (g), and (h),
respectively.  The corresponding $I_{\rm B}$-$V$ characteristics over
the stability range of symmetric breathers are presented in
Fig.~\ref{i-v}(b).

As mentioned above, Figs.~\ref{i-v}(a) and (b) show the $I_{\rm
  B}$-$V$ characteristics with $I_{\rm local}=0$ of asymmetric and
symmetric rotobreathers, respectively.  The voltage $V$ is always
recorded locally on the middle vertical junction which was initially
excited by the local current injection.  The vertical line on the left
side corresponds to the superconducting (static) state.  The rightmost
(also the bottom) curve accounts for the spatially uniform whirling
state (all vertical junctions rotate synchronously and horizontal
junctions are not rotating).  An electrical image of this state is
shown in Fig.~\ref{schematic}(c).  The upper branches in
Fig.~\ref{i-v}
represent various localized states.  The uppermost branch corresponds
to an single-site breather, the next lower branch to a two-site
breather, and so on.

It is easy to show that the voltage $V_{\rm r}$ at which the vertical
and horizontal junctions switch back to the superconducting state is
the same.  By using $R_{\rm sgV}/R_{\rm sgH}=\eta$ and $I_{\rm
  cH}/I_{\rm cV}=\eta$ we get with $I_{\rm r}\propto I_{\rm c}$ that
$V_{\rm r} = R_{\rm sgV} I_{\rm rV} = R_{\rm sgH} I_{\rm rH}$.  For
the uniform state we observed $V_{\rm r}\approx 0.6$ mV.  At about the
same voltage, the whirling horizontal and vertical junctions for the
asymmetric breather are retrapped to the static state as can be seen
in Fig.~\ref{i-v}(a).  The symmetric breathers show different
behavior.  Assuming $V_{\rm H} \approx V_{\rm V}/2$, we can expect
that already at the voltage $V=2 V_{\rm r}\approx 1.2$ mV the
horizontal junctions should trap into the superconducting state. But
if the voltages of the top and bottom horizontal junctions are not
equal while decreasing $I_{\rm B}$, the retrapping current in the
junction with the lower voltage is reached earlier and the retrapping
occurs at a higher measured voltage.

\begin{figure}[htb] 
\vspace{5pt}
\centerline{\epsfig{file=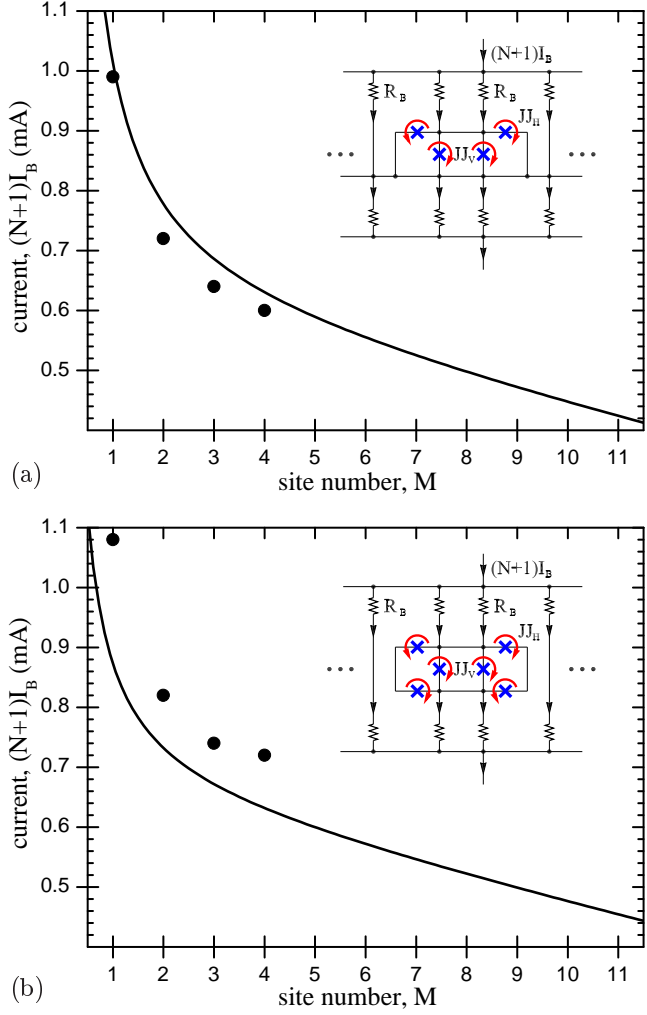,width=3.4in}}
\vspace{5pt}
\caption{The retrapping current measured (points) and calculated
  (line) by Eq.~(\ref{retrapping}) for the asymmetric (a) ($V=V_{\rm
  r}$) and the symmetric (b) ($V=2V_{\rm r}$) rotobreathers.  The
  insets show the reduced electric circuits for each case.}
\label{formula}
\end{figure}

We found that in order to explain the retrapping current of the
breather states the bias resistors $R_{\rm B}$ have to be taken into
account. Initially, these resistors have been designed for providing a
uniform bias current distribution in the ladder. Assuming that the
Josephson junctions in the superconducting state are just short
circuits we get the electric circuits which are shown as insets in
Fig.~\ref{formula}. The total current injected in the breather region
is calculated by using the resistance $R$ of vertical junctions
measured in the spatially-uniform state. Thus we derive the retrapping
current as 
\begin{equation}
  (N+1)I_{\rm B}= \left[\frac{(N-M+1)}{(1+\delta) R_{\rm B}} +
    \frac{(N + 1)[M + (2-\delta) \eta]}{M R}
  \right] V,
\label{retrapping}
\end{equation}
where $M$ is the number of whirling vertical junctions. The parameter
$\delta$ is equal to $0$ for asymmetric breathers and to $1$ for
symmetric breathers. The calculated retrapping current is compared
with the experiment in Fig.~\ref{formula}.  For the asymmetric
breathers the agreement is fairly good. We believe that the larger
discrepancy for the symmetric breathers is due to the assumption
$V_{\rm H} = V_{\rm V}/2$ that we imposed in derivation of the
retrapping current.

\begin{figure}[htb] 
\vspace{5pt}
\centerline{\epsfig{file=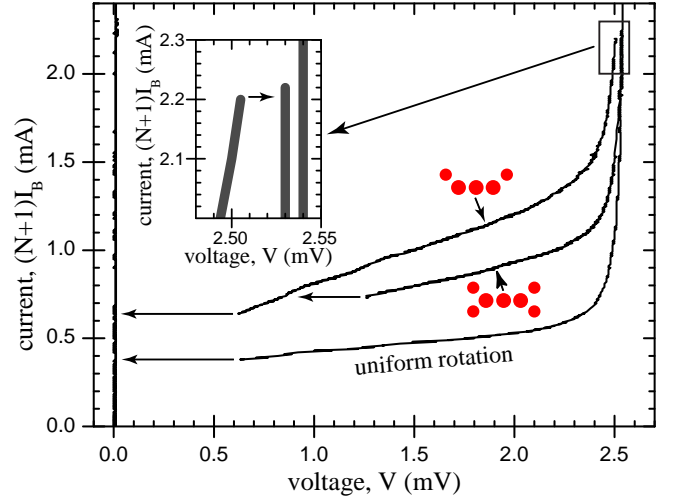,width=3.4in}}
\vspace{5pt}
\caption{$I_{\rm B}$-$V$ characteristics of three-site breathers.}
\label{dynamic}
\end{figure}

One interesting feature in our experiment is the dynamical behavior
of the local states in the linear ladder which differs from the
previous observations in the annular ladder \cite{binder00} with the
anisotropy parameter $\eta = 0.44$.  In
Fig.~\ref{dynamic} this feature is illustrated for 3-site breathers.
After the creation of a symmetric rotobreather (middle curve) we are
lowering the uniform bias current $I_{\rm B}$ until the retrapping
current of one of the horizontal junctions at each side is reached.
Here we observe switching to the corresponding asymmetric 3-site
rotobreather state.  By further lowering of the bias current we get to
the point where both the vertical and the horizontal junctions reach
their retrapping currents and the whole ladder goes into the
superconducting state. If, instead of lowering the bias current, we
start increasing it another switching point is observed where the
previous symmetric breather state is recovered. In contrast to this
behavior, in the annular ladder \cite{binder00} the observed lower
instability of the symmetric breather led to an {\em increase\/} of
voltage. Moreover, this switching turned the number of rotating
junctions to increase, with few more vertical junctions starting to
rotate.  

\begin{figure}[htb]
\vspace{5pt}
\centerline{\epsfig{file=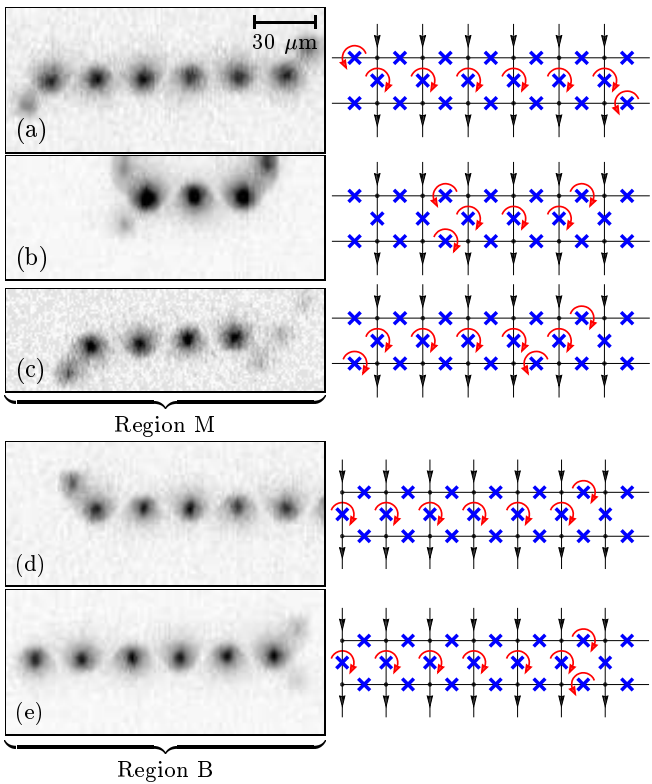,width=3.4in}}
\vspace{5pt}
\caption{More complex non-uniform states measured in the middle (M)
and on the border (B) regions of the ladder.}
\label{sophistic}
\end{figure}

In addition to the simplest hierarchic states described above we also
found a large variety of more complex localized dynamic states.  Some
examples are presented in Fig.~\ref{sophistic}.
Fig.~\ref{sophistic}(a)-(c) are sections from the middle region (M)
[cf.  Fig.~\ref{schematic}(b)] of the measured linear ladder.
Fig.~\ref{sophistic}(a) shows an asymmetric 6-site breather with top
and bottom horizontal junctions whirling on its sides.
Fig.~\ref{sophistic}(b) illustrates a 3-site rotobreather which is
symmetric on one side and asymmetric on the other.  Another very
peculiar localized state is presented in Fig.~\ref{sophistic}(c).  The
rightmost vertical junction is rotating with a lower frequency then
the other four vertical junctions on the left.  This can be understood
from the appearance of an extra rotating horizontal junction in the
interior of this state.  We can interpret it as a 4-site breather
coupled to a single-site breather.  All these localized states would
be topologically forbidden in the case of an annular ladder as the
magnetic flux inside the superconducting circuit should not
accumulate, but they are not forbidden in a ladder with open
boundaries.  The last two pictures (d) and (e) are taken near the
border of the ladder.  We find here truncated asymmetric and symmetric
6-site breathers.  The marginal vertical junction does not require any
horizontal junctions to rotate.

In summary, we presented observation of a large variety of
spatially-localized dynamic rotobreather states in Josephson ladders
with open boundaries. These states can be excited by the local current
injection and supported by the uniform current bias. We observe both
symmetric and asymmetric rotobreather states predicted in
Ref.~\onlinecite{Mazo99}, as well as more complex mixed states. We
believe that the region of existence and stability of rotobreathers
does sensitively depend on dissipation, discreteness, and anisotropy
parameter of the ladder.

\end{document}